\documentclass[journal]{IEEEtran}
\usepackage{amsmath,graphicx}
\usepackage{amssymb,amsfonts,array}
\usepackage{bbm}
\usepackage{bm}
\usepackage{algorithm2e}
\usepackage{cite}
\usepackage{dsfont}
\usepackage{algorithmic}
\usepackage{xcolor}
\usepackage{bbm}
\RestyleAlgo{ruled}
\usepackage{subcaption}

\makeatletter
\def\bstctlcite{\@ifnextchar[{\@bstctlcite}{\@bstctlcite[@auxout]}}
\def\@bstctlcite[#1]#2{\@bsphack
 \@for\@citeb:=#2\do{%
   \edef\@citeb{\expandafter\@firstofone\@citeb}%
   \if@filesw\immediate\write\csname #1\endcsname{\string\citation{\@citeb}}\fi}%
 \@esphack}
\makeatother

\begin{document}
\bstctlcite{BSTcontrol}
\title{A Unified Analysis for Dynamic Programming Track-Before-Detect Algorithms: Error Convergence and Spatial Uncertainty}

\author{Nicholas Bampton, Tian J. Ma, Minh N. Do
\thanks{Received  XX MONTH 202X; revised XX MONTH 202X, XX MONTH 202X,
and XX MONTH 202X; accepted XX MONTH 202X. Date of publication
XX MONTH 202X; date of current version XX MONTH 202X. This work was supported by Sandia National Laboratories. Sandia National Laboratories is a multimission laboratory managed and operated by National Technology \& Engineering Solutions of Sandia, LLC, a wholly owned subsidiary of Honeywell International Inc., for the U.S. Department of Energy’s National Nuclear Security Administration under contract DE-NA0003525. SAND\#0000-XXXXX. This paper describes objective technical results and analysis. Any subjective views or opinions that might be expressed in the paper do not necessarily represent the views of the U.S. Department of Energy or the United States Government.}
\thanks{N. Bampton and M. N. Do are with the University of Illinois at Urbana-Champaign, Champaign, IL 61801 USA (e-mail: bampton2@illinois.edu; minhdo@illinois.edu).}%
\thanks{T. J. Ma is with Sandia National Laboratories, Albuquerque, NM 87185, USA (e-mail: tma@sandia.gov).}%
\thanks{Code is available online at: https://github.com/nick-bampton/Normalized-Path-Integration}}

\maketitle

\begin{abstract}

The Dynamic Programming Track-Before-Detect (DP-TBD) class of algorithms is a core approach to the small low signal-to-noise ratio (SNR) target detection problem. These methods detect targets by recursively accumulating data through a sequence of iterative maximizations, a process that has traditionally limited their theoretical analysis. We propose a novel spatial analysis for the general DP-TBD class of algorithms where we derive a fundamental inverse relationship between detection uncertainty and location uncertainty using specific threshold constructions. Our analysis explicitly incorporates spatial distance from the target state into the probability bounds and allow this distance to vary as a function of iteration count, i.e. the number of processed frames. Integrating additional observations increases confidence in target existence while reducing certainty about the target's location. Our framework precisely details how each parameter affects performance and establishes the necessary conditions under which this analysis holds. Within this framework, we propose Normalized Path Integration (NPI), a DP-TBD algorithm that achieves broad applicability by tracking targets based on the similarity between observations as opposed to directly integrating the observations themselves. We experimentally validate this theory and  compare different DP-TBD constructions on the Sequential Infrared Small Target Detection  (SIRSTD) dataset: a real dataset consisting of  small aerial infrared targets.  

\end{abstract}


\section{Introduction}

\IEEEPARstart{S}{mall} target detection  is a field of remote sensing characterized by foreground elements that are small, low-contrast and featureless.  Many applications arise in long range sensing, including early warning systems, astronomy, and search and rescue, where observations may be subject  to a low signal to noise ratio (SNR). This work is specifically motivated by high altitude infrared small target detection (ISTD), where the frame size can reach the megapixel scale, while target size typically consists of a 1-9 pixels, and target count can be high. These properties necessitate computational efficiency and scalability with the frame size and target count. As small targets are primarily featureless,  the general target detection literature  has limited relevance. Low SNR makes this task particularly challenging as noise can both obscure and mimic the targets; this makes background suppression alone insufficient, complicating detection significantly. 

The target detection and tracking literature is commonly divided into two categories: Detect-Before-Track (DBT) and Track-Before-Detect (TBD). DBT methods identify targets in each frame then associates these detections across frames to create target tracks. TBD  methods \cite{7395388,1541440,4104027} integrate data over multiple frames along possible target trajectories prior to determining target states. Under low SNR conditions, DBT is infeasible as single frame detection is subject to large amounts of  false positives and false negatives. TBD is generally more robust as the integration typically reduces the effect of noise. However these methods suffer from the curse of dimensionality: the number of potential target paths increases exponentially with the number of frames. Consequently TBD methods require various mechanisms and assumptions to achieve computational efficiency; when these mechanisms fail or the assumptions are violated, performance deteriorates and it can potentially render the method entirely inapplicable.

\subsection{Related Work}

The Hough transform\cite{250410,250411} maps frame data into a space of parameterized trajectories, e.g. linear or quadratic motion. Each pixel contributes to the set of trajectories that intersect it, and the local maxima in this parameter space are likely target trajectories. However, including addition motion parameters to represent more complex trajectories quickly increases the computational complexity and memory requirements. 

Dynamic Programming (DP) methods\cite{10128750}, Viterbi \cite{4104027,4104433,6492232}, Generalized Likelihood Ratio (GLR) \cite{249112,6475194} and Pixel Integration \cite{993242,1621242}, accumulate data through recursive maximizations. These algorithms associate real valued weights to each state transition and calculate paths of maximum value to each end state. The dynamic programming implementation is essential to achieve computation efficiency as it reduces exponential number of searched paths. The performance depends on the relationship between target SNR and the state transition size, it is detailed in this paper. 
 
Particle based methods, Particle Filtering (PF)\cite{946220,boers2001particle,9761758,jia2022low} and Sequential Monte Carlo Probability Hypothesis Density  Filtering (SMC-PHD)\cite{1561884}, are derived from Bayesian Filtering where intractable intregrals are numerically approximated using a set of particles.  To maintain estimation accuracy as target SNR decreases, the number of required particles must increase which becomes computationally expensive. In addition the performance of particle methods  depends on the number of targets in the data, as a larger target set necessitates more particles to approximate the relevant integrals.
 
A notable category of TBD are machine learning based methods. Advances in deep learning have resulted in many architectures being adapted for small target detection.  Early approaches were DBT\cite{9423171,10.1109/TIP.2022.3199107,9989433}  as they were adapted from single-frame image classifiers. Recent work has focused on TBD, incorporating temporal information and motion estimation\cite{10381806,10663463}. The performance and generalizability of these methods is largely dictated by the data they are trained on. 

Despite extensive work on temporal integration in TBD, relatively few studies characterize the asymptotic behavior of error probabilities as the integration window increases. Such analysis proves challenging as the target state sequence is unknown and must be inferred jointly with detection. The difficulty primarily stems from the inherent location uncertainty associated with small low SNR target observations.

\subsection{Contribution}

We propose a novel perspective shift in DP-TBD class of algorithms by characterizing the fundamental tradeoff between detection uncertainty and spatial uncertainty. The main result of the paper, Theorem 1  (Section IV.A), proves explicit upper bounds on both the probability of false positives and probability of false negatives. Specifically that under certain threshold constructions as  the number of integrated frames, $k$, increases the probability of misclassifying target states and non-target states decreases exponentially, $O(\exp(-Ak))$, however this bound only holds for non-targets that lie outside a linearly increasing radius from the target, $O(k)$. Hence integrating more data increases the confidence that a target exists, but loses precision about its exact location. We provide a detailed analysis of how key algorithmic parameters influence these bounds, establishing explicit conditions under which DP-TBD methods remain applicable. Within this framework of analysis we introduce a novel DP-TBD algorithm, Normalized Path Integration (NPI); a method that integrates the relationship between sequential observations rather than the individual observations themselves. 

 The paper is organized as follows. Section II defines the general framework for DP-TBD algorithms. Section III details the specific construction for various components within this framework. Section IV derives theoretical analysis for the general framework in Section II and hence provides a unified analysis for all specific DP-TBD algorithms  in Section III. Section V validates the theoretical analysis on a synthetic dataset and compares different DP-TBD constructions on a real infrared dataset. Section VI is the conclusion and discusses future work.

 \begin{figure}[t]
    \centering
\includegraphics[width=0.45\textwidth]{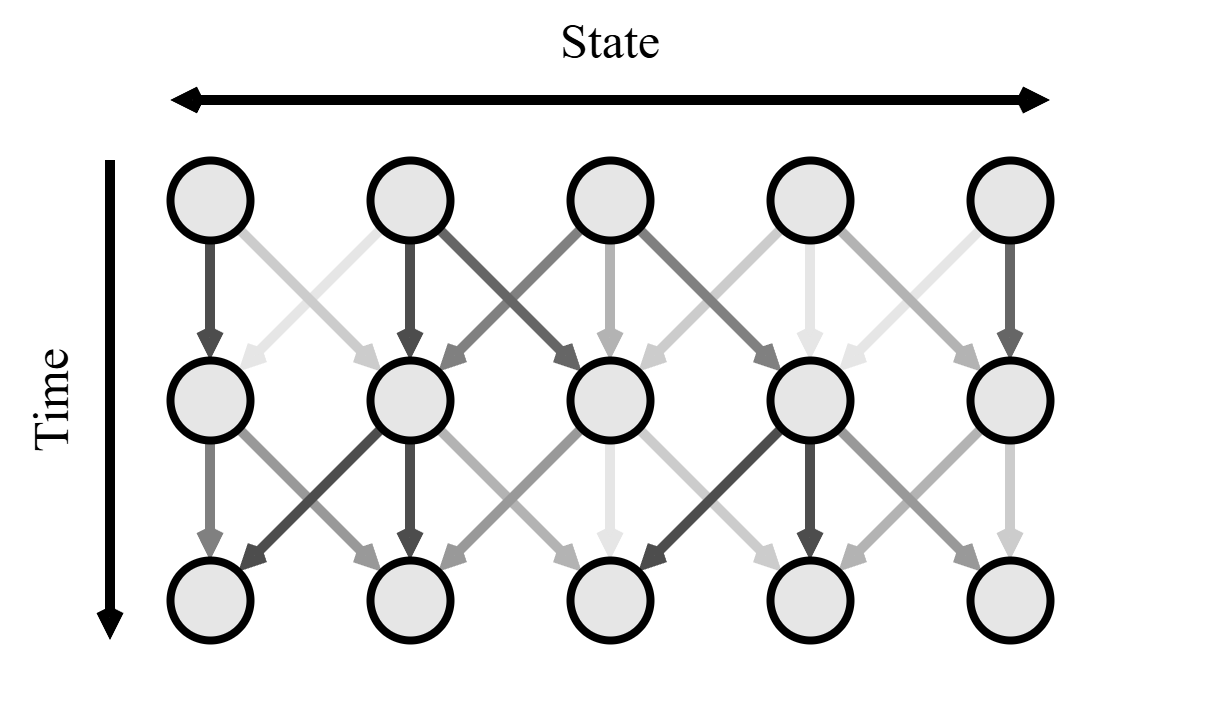}
         \caption{Diagram of $\mathcal{G}_\mathcal{T}$, the graph of all potential target paths. Edge weights are real numbers encoded by the specific DP-TBD algorithm, represented by different shades of gray. }
\end{figure}

\section{General Framework}

In this section we construct the general framework for DP-TBD algorithms. Section II.A defines the graphs structures  and Section II.B defines the general DP-TBD algorithm.

\begin{figure*}[h]
     \centering
     \begin{subfigure}[b]{0.16\textwidth}
         \centering
         \includegraphics[width=\textwidth]{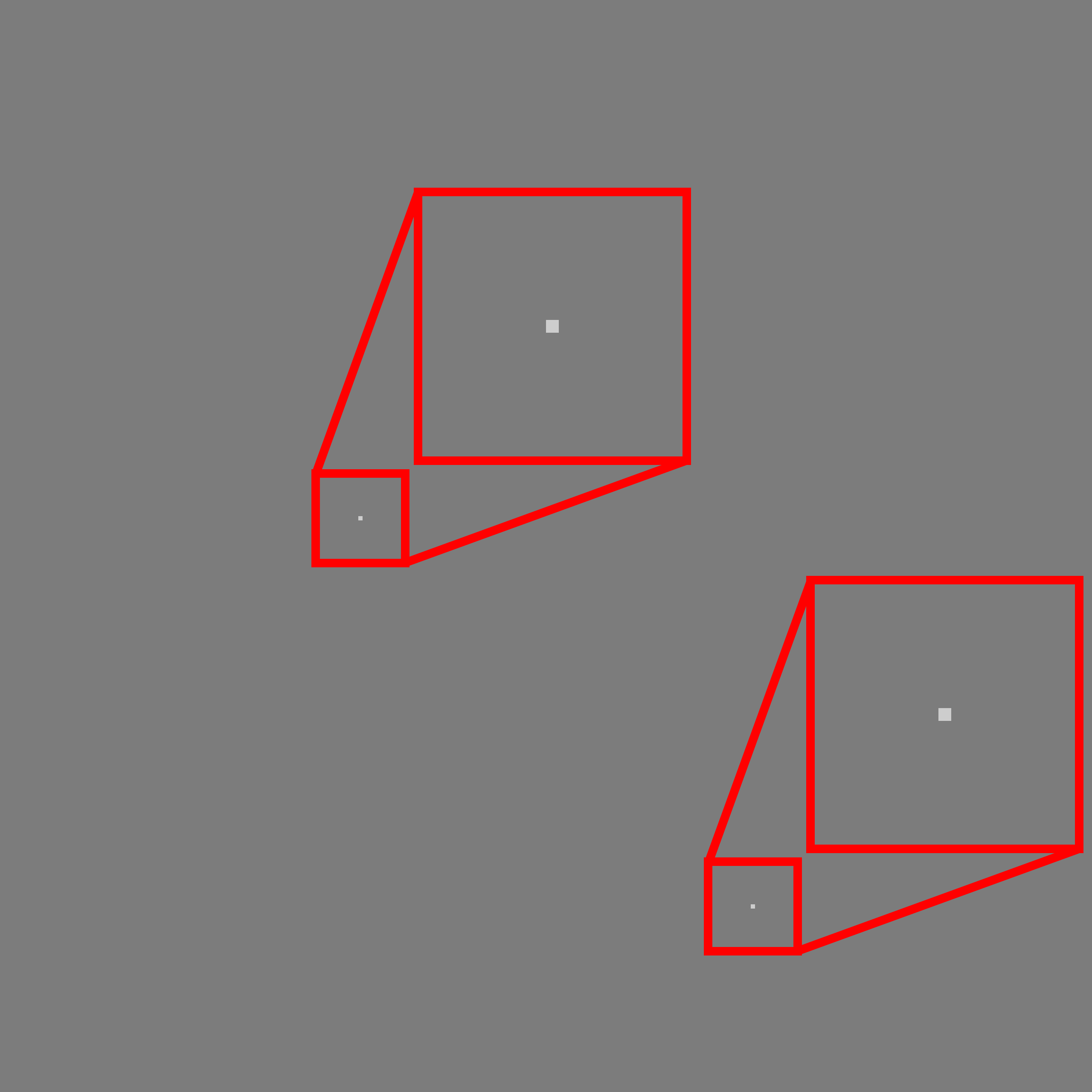}
         \caption{Frame, no noise}
         \label{fig:y equals x}
     \end{subfigure}
     \begin{subfigure}[b]{0.16\textwidth}
         \centering
         \includegraphics[width=\textwidth]{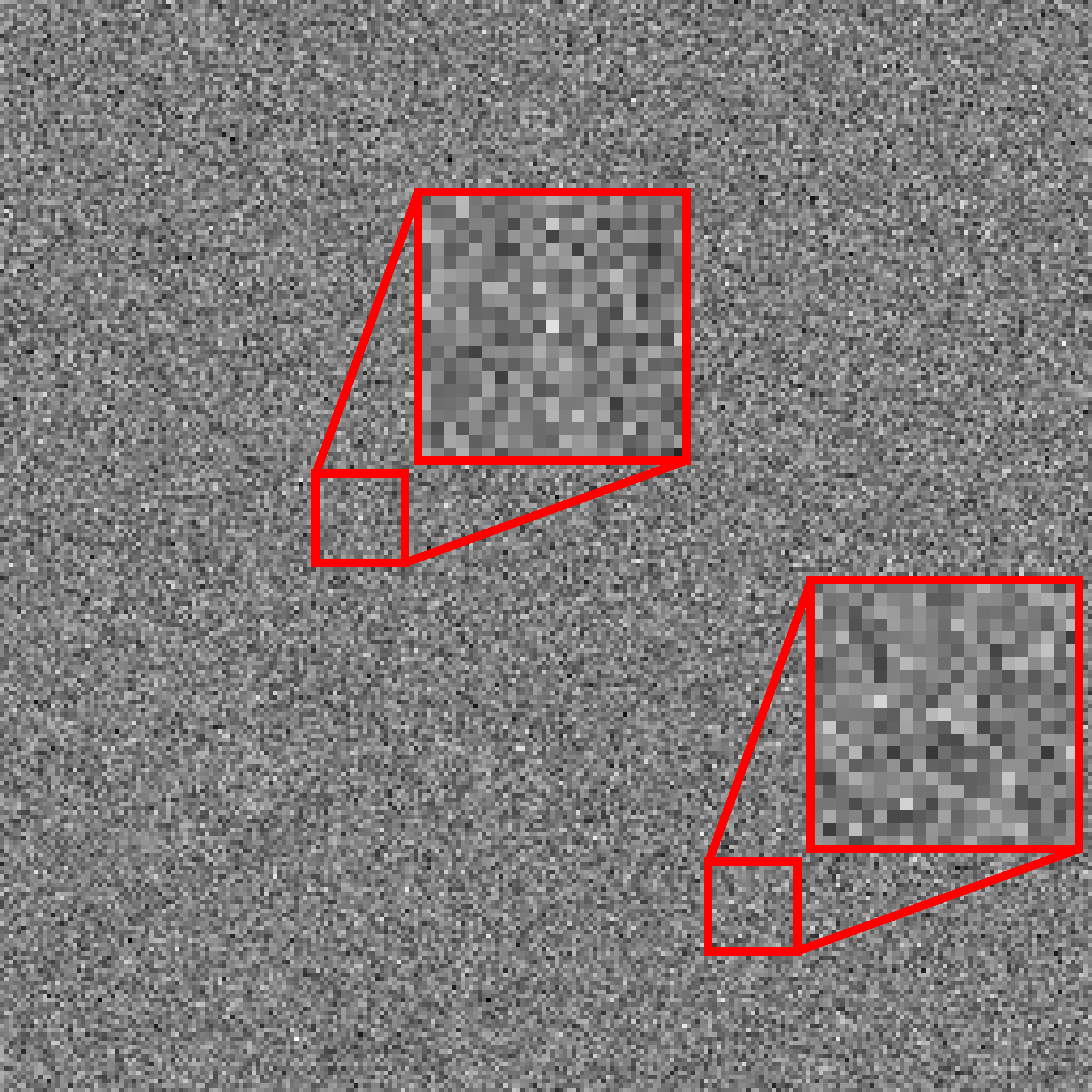}
         \caption{Frame, 2.5 SNR}
         \label{fig:y equals x}
     \end{subfigure}
     \begin{subfigure}[b]{0.16\textwidth}
         \centering
         \includegraphics[width=\textwidth]{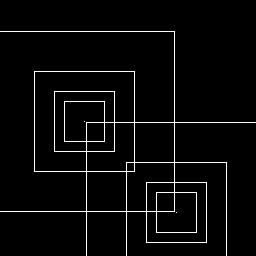}
         \caption{20,30,50,90 radii}
         \label{fig:y equals x}
     \end{subfigure}     
     \begin{subfigure}[b]{0.16\textwidth}
         \centering
         \includegraphics[width=\textwidth]{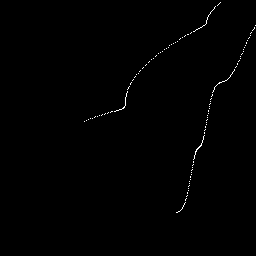}
         \caption{Target Path}
     \end{subfigure}\\
     \centering
     \begin{subfigure}[b]{0.16\textwidth}
         \centering
         \includegraphics[width=\textwidth]{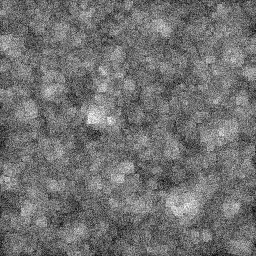}
         \caption{$H_{20}(j,t)$}
         \label{fig:y equals x}
     \end{subfigure}
    \begin{subfigure}[b]{0.16\textwidth}
         \centering
         \includegraphics[width=\textwidth]{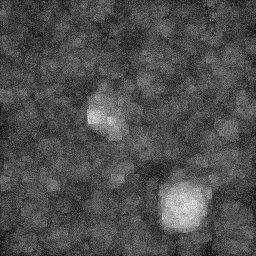}
         \caption{$H_{30}(j,t)$}
         \label{fig:y equals x}
     \end{subfigure}
     \begin{subfigure}[b]{0.16\textwidth}
         \centering
         \includegraphics[width=\textwidth]{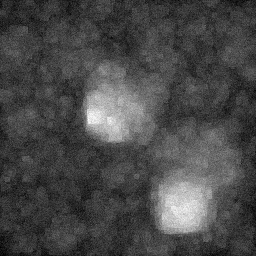}
         \caption{$H_{50}(j,t)$}
         \label{fig:y equals x}
     \end{subfigure}
        \begin{subfigure}[b]{0.16\textwidth}
         \centering
         \includegraphics[width=\textwidth]{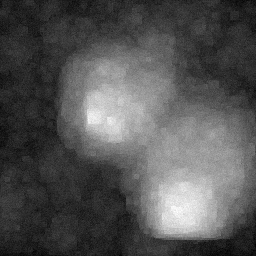}
         \caption{$H_{90}(j,t)$}
         \label{fig:y equals x}
     \end{subfigure}\\
     \centering
     \begin{subfigure}[b]{0.16\textwidth}
         \centering
         \includegraphics[width=\textwidth]{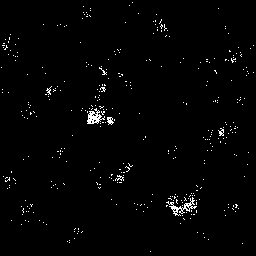}
         \caption{$H_{20}(j,t)>\delta$}
         \label{fig:y equals x}
     \end{subfigure}
    \begin{subfigure}[b]{0.16\textwidth}
         \centering
         \includegraphics[width=\textwidth]{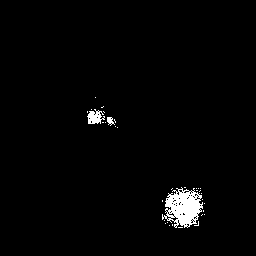}
         \caption{$H_{30}(j,t)>\delta$}
         \label{fig:y equals x}
     \end{subfigure}
     \begin{subfigure}[b]{0.16\textwidth}
         \centering
         \includegraphics[width=\textwidth]{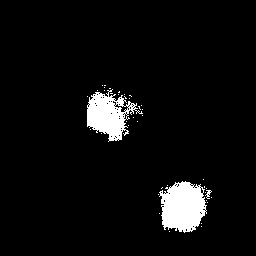}
         \caption{$H_{50}(j,t)>\delta$}
         \label{fig:y equals x}
     \end{subfigure}
        \begin{subfigure}[b]{0.16\textwidth}
         \centering
         \includegraphics[width=\textwidth]{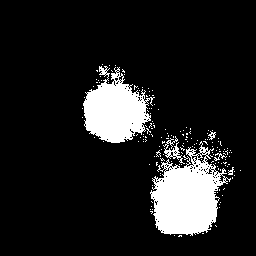}
         \caption{$H_{90}(j,t)>\delta$}
         \label{fig:y equals x}
     \end{subfigure}
        \caption{Effect of integration length $k$ on $H_k(j,t)$ for the PI weight function. (a) is the original uncorrupted frame. (b) is the noisy frame, (c) is the 20,30,50 and 90 pixel radii of the target. (d) is the target trajectory. (e)–(h) are $H_k(j,t)$ for increasing values of $k$ (20, 30, 50, 90).  (i)–(l) are $H_k(j,t)>\delta$ for increasing values of $k$ (20, 30, 50, 90) with $\delta$ from Theorem 2 for $\lambda = 0.7$. Larger $k$ increases target cloud amplitude relative to the background but also increases target cloud width. This results in less false positives and false negatives but a greater radius of the positive regions.}
        \label{fig:three graphs}       
\end{figure*}

\subsection{General Construction}

The standard target tracking model consists of a collection of potential target states, $i\in \mathcal{X}$ (e.g. pixel locations),  and time indexed observations, $Z_t\in \mathcal{Z}$ for $t\in \mathcal{T} = \{0,...,T\}$ (e.g.  sequence of images). Each state has a transition neighborhood, $\mathcal{N}^+_i\subset\mathcal{X}$, the subset of states that a target in state $i$ can transition to in one timestep; the  reverse transition neighborhood is defined $\mathcal{N}^-_i = \{j\in \mathcal{X}: i \in \mathcal{N}^+_j\}$. Additionally we assume a Hidden Markov Model; observations are conditionally independent given a sequence of states.  

We construct a graph $\mathcal{G}(\mathcal{V},\mathcal{E})$ for the purpose of defining  distance within our state space.   It is construced as 
\begin{align}
     \mathcal{V}  &=   \{ i: i\in \mathcal{X}\},\\
     \mathcal{E} &= \{i\rightarrow j  :   j \in \mathcal{N}^+_i\}.
\end{align}

\noindent  This graph induces the shortest path metric, $d_\mathcal{G}(i,j)$, defined as the minimum number of edges required to connect state $i$ to state $j$. If $d_\mathcal{G}(i,j)\leq m$, then a target in state $i$ can reach state $j$ within  $m$ time steps. We propagate this graph along the time dimension,  creating a new graph, $\mathcal{G}_\mathcal{T}(\mathcal{V}_\mathcal{T},\mathcal{E}_\mathcal{T})$. This graph serves as the underlying structure for our analysis, 
\begin{align}
     \mathcal{V}_\mathcal{T}  &=   \{ (i,t) : i \in V, t\in  \mathcal{T} \},\\
     \mathcal{E}_\mathcal{T} &= \{(i,t-1)\rightarrow (j,t) :  i\rightarrow j \in E\}.
\end{align}

\noindent $\mathcal{E}_\mathcal{T}$ is the set of all target state transitions at all time steps therefore every possible target trajectory exists as a path in $\mathcal{G}_\mathcal{T}$. The observations are encoded as edge weights in this graph. We denote $W_t^{ij}\in \mathbb{R}$ as the weight of edge $(i,t-1)\rightarrow (j,t)$; it is constructed as a function of the following form
 \begin{equation}
    W_t^{ij} = f(Z_{t-1},Z_t,i,j).
 \end{equation}

\noindent The specific function is determined by the specific DP-TBD algorithm. Not every DP-TBD methods use all inputs, GLR and Viterbi  \cite{4104027,4104433,249112,6492232} use $f(Z_t,i,j)$,  Pixel Integration \cite{993242,1621242,6475194} uses $f(Z_t,j)$. The required properties of edge weights, (A1)-(A4),  are discussed in Section IV.A.

\subsection{General Algorithm}

Within this framework, all DP-TBD algorithms can be expressed as finding the longest $k$-length path (LP-$k$) to each vertex. Although, for analysis we use the longest $k$-length path average (LPA-$k$). To define the LPA-$k$, we first need to define a path average. Let $\overrightarrow{x} = \{  (x_{t-k},t-k),...,(x_t,t)  \}$ be any path in $\mathcal{G}_\mathcal{T}$, the path average of $\overrightarrow{x}$ is then
\begin{equation}
h_k(\overrightarrow{x})   =    \frac{1}{k}  \textstyle \sum_{\tau =t-k+1}^t   W_\tau^{x_{\tau-1},x_\tau},
\end{equation}

\noindent the average of the edge weights along that path. We define the set of all $k$-length paths to a given vertex, $(j,t)$, as
\begin{equation}
S_k(j,t)  =  \{  \overrightarrow{x} \subset \mathcal{V}_\mathcal{T}     :     \overrightarrow{x}  \  \text{is a $k$-length path to} \ (j,t)   \}.
\end{equation}

\noindent Which lets us formally define the LPA-$k$ to that vertex,
\begin{equation}
      H_{k}(j,t) =  \text{max}_{  \overrightarrow{x}\in S_k(j,t)  } \, h_k(\overrightarrow{x}).
\end{equation}

\noindent  The LPA-$k$ can be computed with the following recursive algorithm and efficiently implemented with dynamic programming,
\begin{align}
     F_{k}(j,t)  &=   \begin{cases}
			\text{max}_{  i\in \mathcal{N}^-_j  } \, W^{ij}_t  + F_{k-1}(i,t-1) & \text{if $k>0$}\\
            0  & \text{if $k=0$}  
            \end{cases}, \notag\\
            H_k(j,t)  &= \frac{1}{k} F_k(j,t)  ,
\end{align} 

\noindent where $F_k(j,t)$ is the LP-$k$ to vertex $(j,t)$. We use the formal definition, (8), in our theoretical derivation and the recursive definition, (9), to  compute $H_k(j,t)$ in practice.

\section{Specific Constructions}

In this section we construct components of the general framework in Section II. Sections III.A-B are state spaces: position and position-velocity. Section III.C is the observation space for grayscale image sequences. Sections III.D-G are the edge weight functions of different DP-TBD algorithms: Pixel Integration (PI), Viterbi, Generalized Likelihood Ratio (GLR), and Normalized Path Integration (NPI).

\subsection{State Space: Position}

The fundamental state space in target tracking consists of   the potential target positions. It is the smallest possible state space and as a result is the most computationally efficient. This is the only applicable state space when target velocity can change sufficiently fast in the time between observations. 
\begin{align}
\mathcal{X}  &=  \{(i_1,i_2)\in \mathbb{N}^2 : i_1<N_1,i_2<N_2\},
\end{align}

\noindent where $N_1$ and $N_2$ are the frame dimensions. To define state transitions we require  target velocity be bounded by some $v_\text{max}$.
\begin{align}
\mathcal{N}^+_i =\{j\in \mathcal{X}:||(i_1,i_2)-(j_1,j_2)||_2\leq v_{\text{max}}\}.
\end{align}

The primary issue when increasing $v_{\text{max}}$ to track faster targets is that $d_\mathcal{G}(x_t,y_t)>m$  equates to $||x_t-y_t||_2>mv_{\text{max}}$; converting from graph distance to Euclidean distance is linear with $v_{\text{max}}$. It also increases the transition neighborhoods size resulting in worse performance, Theorem 1 (Section IV.A).

\subsection{State Space: Position-Velocity}

Incorporating velocity into the state space can increase  performance, as the added information can be used to further limit the size of transition neighborhoods. However, by introducing a new dimension, we increase the size of the state space and therefore the computational complexity. 
\begin{align}
\mathcal{X}  =  \{(i_1,i_2,i_3,i_4)\in \mathbb{N}^2 \times \mathbb{Z}^2 :{}& i_1<N_1, i_2<N_2,\\
&|i_3| < v_\text{max}, |i_4|<v_\text{max}\},\notag
\end{align}

\noindent where $i_1,i_2$ are the position arguments and $i_3,i_4$ are the velocity arguments. To define state transitions we require target acceleration be bounded by some $a_\text{max}$.  
\begin{align}
\mathcal{N}^+_i =\{j\in \mathcal{X} :{}&|| (i_3,i_4) - (j_3,j_4)||_\mathcal{X}<a_\text{max}, \\ 
&  (j_1,j_2) = (i_1,i_2)+(j_3,j_4) \}.\notag
\end{align}

Often the primarily concern is determining target location, in these instances we compute an additional  maximum over all velocities, $\text{max}_{j_3,j_4} H_k((j_1,j_2,j_3,j_4),t)$. We can continue this process of incorporating higher derivatives into the state space, however, this is uncommon as any potential performance benefit does not justify the increase in computation.

\subsection{Observation Space: Image Sequence}

For a grayscale image sequence, each observation, $Z_t$, is  the corresponding input frame, $I_t\in \mathbb{R}^{N_1 \times N_2}$. We may localize these observations by dividing the frame into the set of overlapping square windows, $Z_t^i$, which we refer to as a state observation; for a position state $i=(i_1,i_2)$, or position-velocity state $i = (i_1,i_2,i_3,i_4)$ this is defined as
\begin{align}
Z_t^i &= I_t[i_1-r:i_1+r,i_2-r:i_2+r],
\end{align}

\noindent where $r\geq 0$ is some chosen integer. When $r=0$ state observations are pixels. If the data has Gaussian noise then we can  model the state observations as Gaussian random vectors,
  \begin{align}
 Z_t^i & \sim \mathcal{N}(\mu_t^i,\sigma^2I),
 \end{align}

\noindent where $\mu_t^i$ is the vector corresponding to the section of target and background in the window centered at pixel $i$ at time $t$ and $\sigma^2$ is the Gaussian noise variance.   

\subsection{Weight Function: State Integration}

The most intuitive DP-TBD algorithm is to directly integrate state observations. Edge weights are constructed as the average of the state observation which corresponds to the destination vertex of that edge. When $r=0$, this corresponds to integrating pixel values along trajectories, i.e. pixel integration (PI),
  \begin{align}
 W_t^{ij} & = \text{mean}(Z_{t}^j),
 \end{align}

\noindent  Depending on the data we may instead use the absolute value of the state observation, $W_t^{ij} = \text{mean$(|Z_t^j|)$}$, denoted PI-abs.

PI is particularly appropriate for ISTD as targets are assumed to have higher amplitude than the background. PI retains sign information, that is lost when using PI-abs; this results in higher performance but the inability to track negative amplitude targets.

\subsection{Weight Function: Viterbi}

The Viterbi algorithm finds the sequence of states, $y_{0:t}$, with maximum probability given a sequence of observations, $Z_{1:t}$; although, we are primarily concerned with the final state as it represents the current target location. It is constructed using
\begin{align}
V(j,t) &=  \text{max}_{  \overrightarrow{x}\in S_t(j,t)  } \, P(  x_{0:t} | Z_{1:t})
\end{align}

\noindent The Markov assumptions allow  the following decomposition,
\begin{align}
P(  x_{0:t} |  Z_{1:t})=  \frac{P(x_0)}{P(Z_{t:t})}\textstyle\prod_{\tau=1}^t P(  x_t | x_{1:t-1})P(Z_t|x_t).
\end{align}

\noindent We assume $P(x_0)$ to be uniformly distributed over the state space, so $P(x_0)/P(Z_{1:t}) = K$ is constant. Then we use  $\ln$ to convert between multiplication and summation, 
\begin{align}
W_t^{ij} &= \ln(P(x_t = j | x_{t-1} = i)P(Z_{t}|x_t = j)).
\end{align}

We can recover $V(j,t) = K\exp(tH_t(j,t))$; however, for the purpose of analysis it is sufficient to leave it as $H_t(j,t)$ due to the monotonicity property of $\ln$. Note that traditionally Viterbi does not use a fixed $k$, it continually iterates as each new observation is received.

\subsection{Weight Function:  Generalized Likelihood Ratio}

Generalized Likelihood Ratio (GLR) is a more practical version of Viterbi. Given a sequence of states, the GLR is the probability that the sequence was generated by a target divided by the probability that it was generated by noise
\begin{align}
V(j,t) &= \text{max}_{  \overrightarrow{x}\in S_t(j,t)  } \, \frac{P(  x_{0:t} | Z_{1:t}, H_1)}{P(  x_{0:t} | Z_{1:t}, H_0)}.
\end{align}

\noindent where $H_1$ is the target hypothesis and $H_0$ is the null hypothesis. The numerator and denominator decompose as in (20) 
\begin{align}
W_t^{ij} &= \ln(P(x_t = j | x_{t-1} = i,H_1)P(Z_{t}|x_t = j,H_1))\\
&\color{white}=\color{black}-\ln(P(x_t = j \, | \, x_{t-1} = i,H_0)P(Z_{t} |x_t = j,H_0))\notag
\end{align}
 
For $r=0$ in (14),  if the target transition probabilities are uniform over the transition neighborhood and the target amplitude is assumed to be constant then this is equivalent to PI times a fixed constant. If the target amplitude is unknown then this is equivalent to  $W_t^{ij}  \propto (Z_{t}^j)^2$. We use the latter for experimental validation and numerical results.

\subsection{Weight Function: Normalized Path Integration}

We propose Normalized Path Integration (NPI), it uses the similarity between observations rather than the observations themselves. Target state observations are similar to one another and dissimilar from background state observations; normalizing over the transition neighborhood generates a signal
\begin{align}
W_t^{ij} &  = \frac{\epsilon+\exp(-b||Z_t^j-Z_{t-1}^i||_\mathcal{Z}^2)}{\textstyle\sum_{k\in \mathcal{N}^+_i}(\epsilon+\exp(-b||Z_t^k-Z_{t-1}^i||_\mathcal{Z}^2))},
\end{align}
 
\noindent where $b>0$ is optimized for performance and $\epsilon>0$  acts as a regularizer that notably makes this construction Lipschitz.  

NPI is designed to be generally applicable; it does not require known universal target features to construct edge weights which is useful in more complex observation spaces (e.g. rgb image sequences). It is  robust to certain background changes and has application in correcting for background motion; discussed in Section V.D and Section VII.A.

\section{Analyses}

In this section we analyze the performance of the general DP-TBD algorithm construction. Section IV.A derives the main theoretical results. Section IV.B proves this theory  for practical implementation. Sections IV.C discusses background signals.  Section IV.D discusses the time and space complexity of these algorithms. All proofs for lemmas and theorems are in the Appendix.

\subsection{Theoretical Analysis}

After computing $H_k(j, t)$, we select a decision threshold, $\delta$,  where a vertex $(j, t)$ is classified positive if $H_k(j, t) > \delta$, and negative if $H_k(j, t) < \delta$.  In this section we derive bounds for the probability of false positives and false negatives as a function of path length, $k$. For simplicity consider a single target where  $y_t$ denotes the target state at time $t$. We require the following assumptions. 
\begin{alignat*}{2}
\text{(A1)} \ \ \  &\mathbb{E}[  W_t^{ij}   ] \geq \mu_1 \quad   \forall i,j,t \;\, \text{where} \;\, i = y_{t-1}, \;\, j = y_{t}, \notag \\
\text{(A2)} \ \ \  &\mathbb{E}[  W_t^{ij}  ]\leq \mu_2       \quad \forall i,j,t \;\, \text{where}  \;\, j \neq y_{t}, \notag \\
\text{(A3)} \ \ \  &|  \mathcal{N}^-_i  |\leq M \quad\forall i. \color{white}W_t^{ij}\color{black} \notag\\
 \text{(A4)} \ \ \ &  W_t^{ij} - \mathbb{E}[  W_t^{ij}  ]   \text{ is $C$-subgaussian} \quad\forall i,j,t\notag.
\end{alignat*}

\noindent (A1) lower bounds the expected value of target edge weights, while (A2) upper bounds the expected value of non-target edge weights.  (A3) bounds the size of predecessor neighborhoods. (A4) ensures  the tail probabilities of edge weights decay at least as fast as those of a Gaussian with variance $C^2$. An important result is that a Lipschitz function of Gaussian random variables is subgaussian \cite[Theorem 2.26]{wainwright2019high}. Additionally all derivations can be further generalized with an analogous subexponential definition. In isolation, these assumptions are relatively common;  however, the applicability of this theory is determined by the relationship between $\mu_1$, $\mu_2$, $M$, and $C$. \vspace*{6pt}

\noindent \textit{Lemma 1.1:} Let $\overrightarrow{y} \in S_k(y_t,t)$ be the $k$-length target path and $\overrightarrow{x} \in S_k(x_t,t)$ be any $k$-length path where $d_\mathcal{G}(x_t,y_t)>m$, for $m\in(0,2k]$. If $\mu_1>\mu_2$,  then there exists $ \alpha\in \mathbb{R}$ such that 
\begin{subequations}
\begin{align}
\mathbb{E}[  h_k( \overrightarrow{y} )  ]  &\geq   \alpha + \frac{m}{2k}\mu_1,\\
\mathbb{E}[  h_k( \overrightarrow{x} )  ]  &<   \alpha + \frac{m}{2k}\mu_2. 
\end{align}
\end{subequations}

\noindent This lemma follows directly from the triangle inequality; $d_\mathcal{G}(x_t,y_t)\leq d_\mathcal{G}(x_t,y_{t-m/2}) +  d_\mathcal{G}(y_t,y_{t-m/2})$ implies that  $d_\mathcal{G}(x_t,y_{t-m/2})>m/2$ which means the last $m/2$ edges in any path to $(x_t,t)$ cannot be target edges. The shared term, $\alpha$,  allows us to relate the  false positive and false negative probabilities.   \vspace*{6pt} 

\noindent \textit{Lemma 1.2:} Let $\overrightarrow{x}$ be a $k$-length path in $\mathcal{G}_\mathcal{T}$. If (A4) is satisfied then $h_k(\overrightarrow{x}) - \mathbb{E}[  h_k(\overrightarrow{x})  ]$ is  $\sqrt{2C^2}/k$-subgaussian.\vspace*{6pt}

\noindent   The derivation is similar to the proof of the law of large numbers:  the subgaussian constant, $C$, behaves similarly to variance when adding random variables. Although there are minor adjustments to address the potential dependency between sequential edge weights. When  all $W_t^{ij}$ are independent this lemma simplifies to $C/k$-subgaussian.

\noindent \textit{Theorem 1:}  Let $\rho \in (0,1]$ and $y_t$ be the target state.  If $\rho >  2C\sqrt{  \ln(M)} / (\mu_1-\mu_2)$,    then there exists $A\in(0,\infty)$, independent of $k$ and $t$, and $ \delta\in\mathbb{R}$ such that for  any non-target state $x_t$  satisfying  $d_\mathcal{G}(x_t,y_t)>2\rho k$ 
\begin{subequations}
\begin{align}
P(  H_k(y_t,t)   <   \delta  )   &<   \exp(  -Ak  ),\\
P(  H_k(x_t,t)   >   \delta  )   &<  \exp( -Ak  ).
\end{align}
\end{subequations}

\noindent Theorem 1 bounds the probability of classification error for the target state and for any non-target state sufficiently far from the target; we call this distance the uncertainty radius. Increasing the path length, $k$, results in i) an exponential decrease, $O(\exp(-Ak))$, in the error  bounds, and ii) a linear increase, $O(k)$, in the uncertainty radius. We visualize this in Fig. 2. and numerically validate it in Fig. 4. The uncertainty radius can be decreased to $d_\mathcal{G}(x_t,y_{t-\rho k})>\rho k$ but note the regions center changes to $y_{t-\rho k}$.

The applicability condition, $\rho >  2C\sqrt{  \ln(M)}/(\mu_1-\mu_2)$, has an intuitive interpretation as a comparison between the target path, $\overrightarrow{y}$, and the maximum of non-target paths, $\overrightarrow{x}\in S_k(x_t,t) $,
\begin{equation}
\mathbb{E}[h_k(\overrightarrow{y})] > \mathbb{E}[H_k(x_t,t) ] \;\, \text{for} \;\, d_\mathcal{G}(x_t,y_t)>2 k.
\end{equation}

\noindent This holds for all $k\geq 1$, using $k=1$ or $k=2$ can give a computationally practical heuristic as an alternative to the applicability condition. Put simply, integrating the target signal must be greater than integrating noise. It derives from $\rho = 1$,  where (A1) results in $\mathbb{E}[h_k(\overrightarrow{y})]\geq \mu_1$ and  (A2)-(A4) result in $\mu_2 +2C\sqrt{  \ln(M)} > \mathbb{E}[H_k(x_t,t) ]$ for $d_\mathcal{G}(x_t,y_t)>2 k$. 

The generalized notion of SNR for DP-TBD algorithms is the  target signal amplitude divided by the noise standard deviation:  $\text{SNR} = (\mu_1-\mu_2)/\sigma$. A property of subgaussian distributions is that $C\geq \sigma$, which results in $\rho>2\sqrt{\ln(M)} / \text{SNR}$. Furthermore, allowing  $\rho=1$,  reduces the inequality to 
\begin{equation}
\text{SNR}>2\sqrt{\ln(M)}.
\end{equation}

\noindent Note that $2\sqrt{\ln(M)}$ decreases to $\sqrt{2\ln(M)}$ when edge weights are a function of only one observation.  For PI with a point target of constant amplitude, $s$, and Gaussian noise, the generalized SNR is equivalent to the traditional SNR of a point target, $\text{SNR} = s/\sigma$.

\begin{figure*}[h]
     \centering
     \begin{subfigure}[b]{0.48\textwidth}
         \centering
         \includegraphics[scale=0.45]{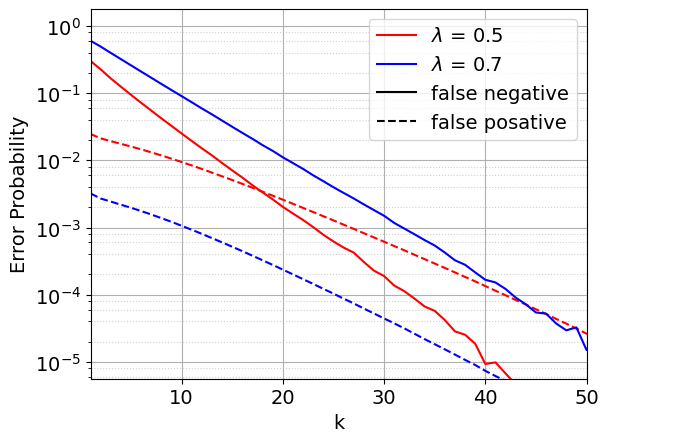} 
         \caption{PI}
         \label{fig:y equals x}
     \end{subfigure}
     \begin{subfigure}[b]{0.48\textwidth}
         \centering
         \includegraphics[scale=0.45]{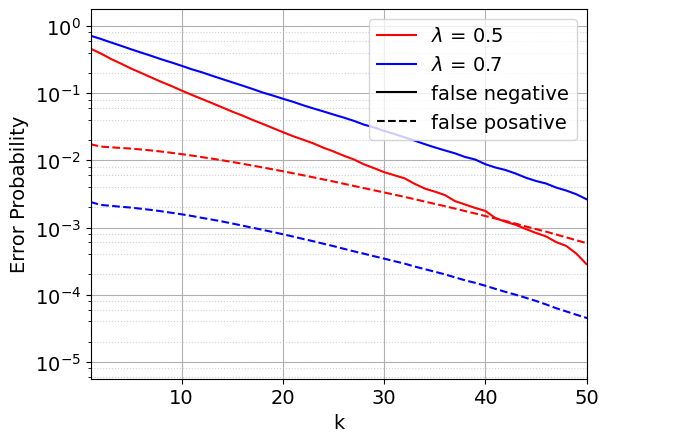}
         \caption{PI-abs}
         \label{fig:y equals x}
     \end{subfigure}\\
          \begin{subfigure}[b]{0.48\textwidth}
         \centering
         \includegraphics[scale=0.45]{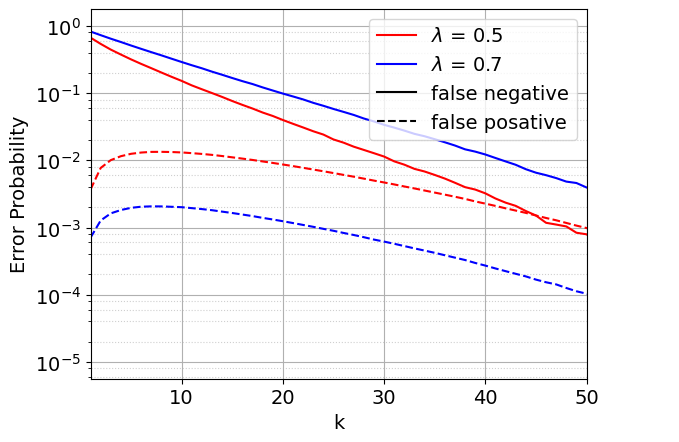}
         \caption{GLR}
         \label{fig:y equals x}
     \end{subfigure}
     \begin{subfigure}[b]{0.48\textwidth}
         \centering
        \includegraphics[scale=0.45]{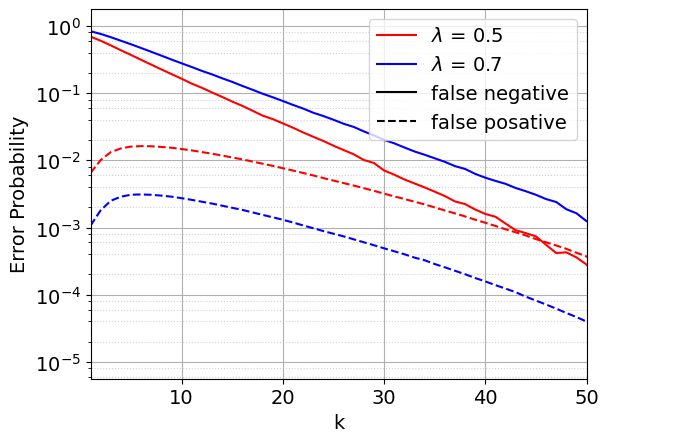}
         \caption{NPI}
         \label{fig:y equals x}
     \end{subfigure}
     \caption{The false positive  and false negative probabilities plotted against the number of integrated frames, $k$, for different DP-TBD methods. The threshold is computed with (27). The false positive rate is measured outside a $k$ pixel radius from the target (see Fig. 2. for visualization). The y-axis is $\log$ scale so  negative linear functions correspond to exponential convergence. }
\end{figure*}

\subsection{Practical Analysis}

In Theorem 1, we choose $\rho$ and $k$ to obtain $A$ and $\delta$, an appropriate order for observing the relationship between different aspects of performance. However, $\delta$  is computed using a priori knowledge and is dependent on $t$. We derive a threshold construction that preserves the relationship from Theorem 1 however we choose  $\delta$ and $k$ to obtain $A$ and $\rho$.  
\vspace*{6pt}

\noindent \textit{Lemma 2.1:}  Let $\delta$ be the Theorem 1 threshold for  $\rho$, $k$, $t$ and let $y_t$ be the target state. For any  non-target state $x_t$ satisfying $d_\mathcal{G}(x_t,y_t)>2 k$ there exists $ \lambda\in [0,1]$  such that 
\begin{equation}
   \delta =  \lambda\mathbb{E}[  H_k(y_t,t)   ] + (1-\lambda)\mathbb{E}[  H_k(x_t,t)   ].
\end{equation}

\noindent This construction lets us choose $\lambda$ as an equivalent to choosing $\delta$. Both $\mathbb{E}[  H_k(y_t,t)   ]$ and $\mathbb{E}[  H_k(x_t,t)   ]$ can be directly approximated  at each time step using $H_k(j,t)$.  We  separate $\mathcal{X}$ into target segments and the background segment. A simplistic method is to threshold $H_k(j,t)$ with a chosen lower detection limit; positive regions are target segments, the negative states are the background segment. Let $U_1\subset\mathcal{X}$ be the  target segment and $U_2\subset\mathcal{X}$ be the background segment. We approximate the values in (27) as
\begin{align}
\mathbb{E}[  H_k(y_t,t)   ]  &\approx   \text{max}_{  j\in U_1  } \, H_k(j,t) ,\\
\mathbb{E}[  H_k(x_t,t)   ]  &\approx   \text{mean}_{  j\in U_2  } \, H_k(j,t).
\end{align}

\noindent This construction naturally extends to multiple target detection;  a separate threshold is defined for each target cloud and applied solely to its corresponding states.  \vspace*{6pt} 

\noindent \textit{Theorem 2:}  Let  $\lambda \in [\beta_0,\beta_1]$ with corresponding threshold $\delta$ from (27)  and let $y_t$ be the target state. Then there exists $A\in(0,\infty)$ and $\rho\in (0,1]$, both independent of $k$ and $t$, such that for any non-target state $x_t$ satisfying $d_\mathcal{G}(x_t,y_t)>2\rho k$ 
\begin{subequations}
\begin{align}
P(  H_k(y_t,t)   <   \delta  )   &<   \exp(  -Ak  ),\\
P(  H_k(x_t,t)   >   \delta  )   &<  \exp( -Ak  ).
\end{align}
\end{subequations}

\noindent This formally proves that the threshold construction in (27) with a fixed $\lambda$   preserves the inverse relationship derived in Theorem 1. The constants $\beta_0,\beta_1$ are defined by the data; if target SNR is constant (i.e. $\mathbb{E}[W_t^{ij}| i = y_{t-1}, j=y_t] = \mu_1$) then $\beta_0 \approx 1/2$ and $\beta_1 \approx 1$. It becomes significantly more complex when target SNR is variable but generally a good choice is $\lambda = 0.7$.

\begin{figure*}[t]
     \centering
     \begin{subfigure}[b]{0.16\textwidth}
         \centering
         \includegraphics[width=\textwidth]{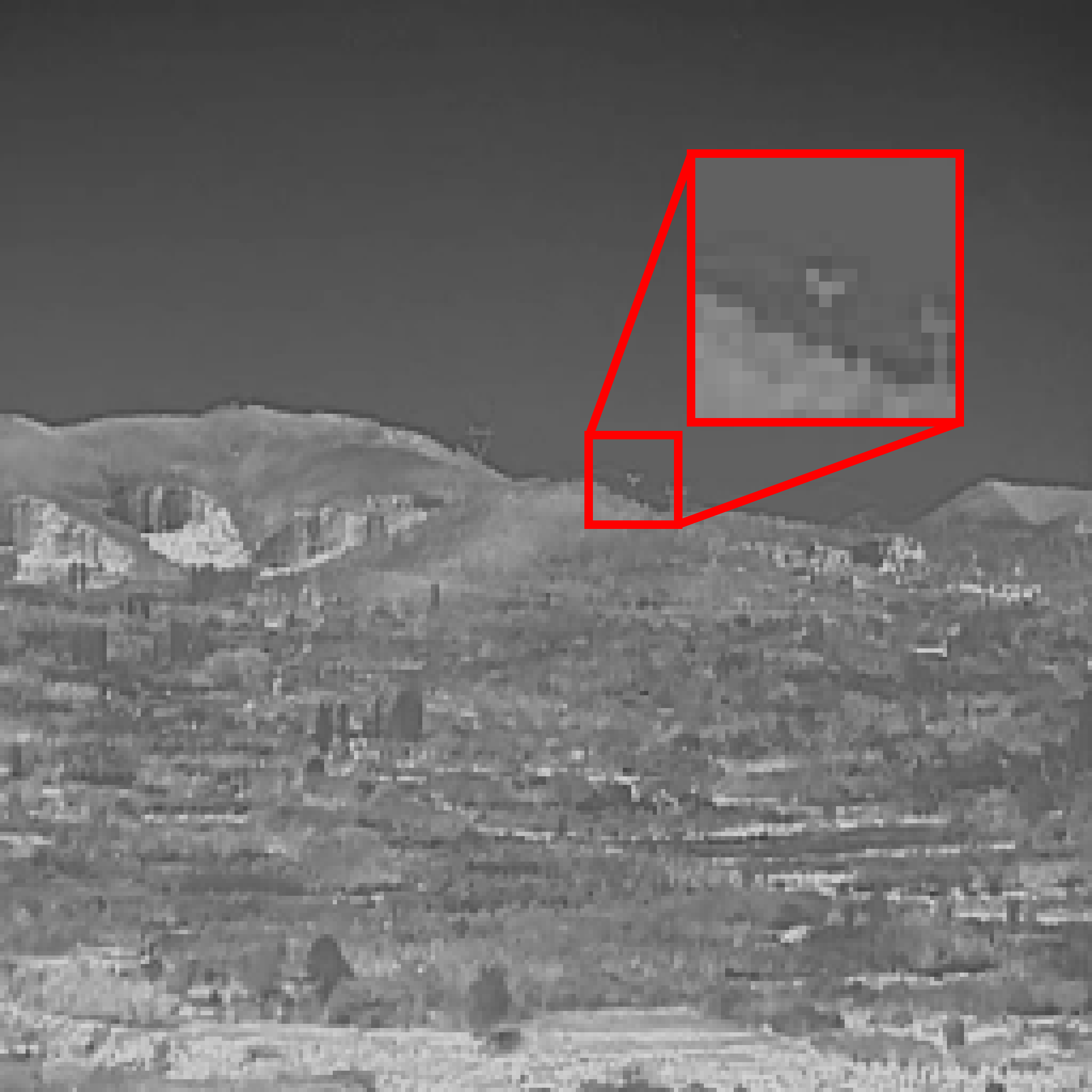}
         \caption{Frame, No Noise}
         \label{fig:y equals x}
     \end{subfigure}
     \begin{subfigure}[b]{0.16\textwidth}
         \centering
         \includegraphics[width=\textwidth]{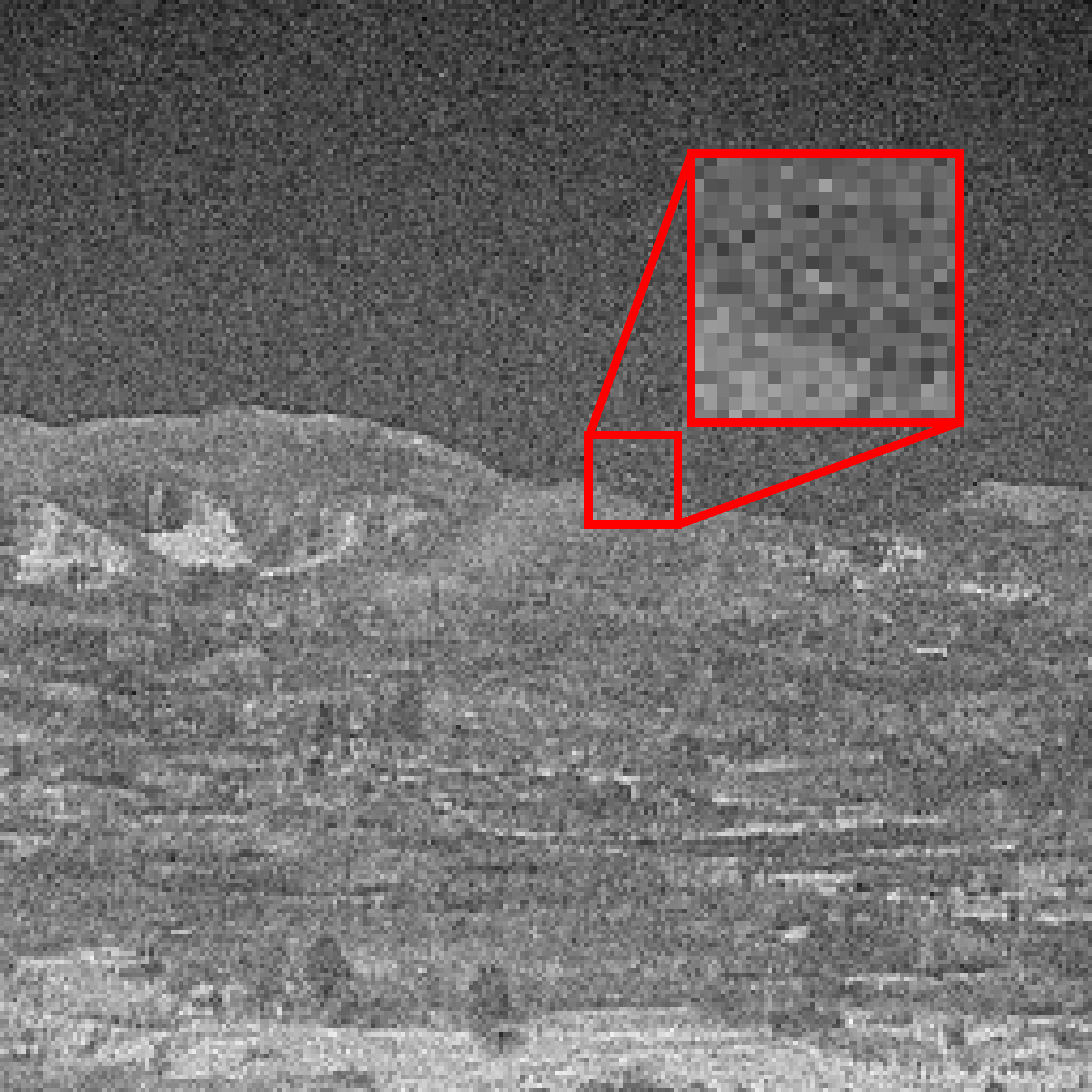}
         \caption{Frame, 2.5 SNR}
         \label{fig:y equals x}
     \end{subfigure}
     \begin{subfigure}[b]{0.16\textwidth}
         \centering
         \includegraphics[width=\textwidth]{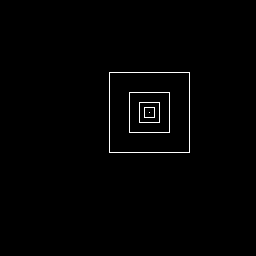}
         \caption{5,10,20,40 Radii}
         \label{fig:y equals x}
     \end{subfigure}
     \begin{subfigure}[b]{0.16\textwidth}
         \centering
         \includegraphics[width=\textwidth]{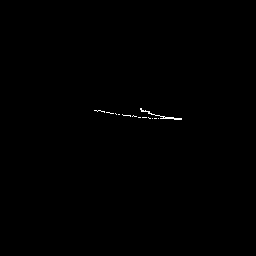}
         \caption{Target Path}
         \label{fig:y equals x}
     \end{subfigure} \\

      \begin{subfigure}[b]{0.16\textwidth}
         \centering
         \includegraphics[width=\textwidth]{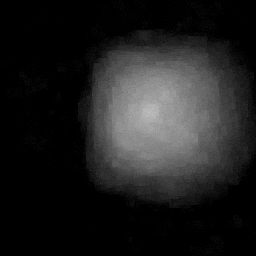}
         \caption{PI Pos}
         \label{fig:y equals x}
     \end{subfigure}
     \begin{subfigure}[b]{0.16\textwidth}
         \centering
         \includegraphics[width=\textwidth]{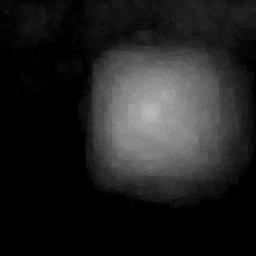}
         \caption{GLR Pos}
         \label{fig:y equals x}
     \end{subfigure}
     \begin{subfigure}[b]{0.16\textwidth}
         \centering
         \includegraphics[width=\textwidth]{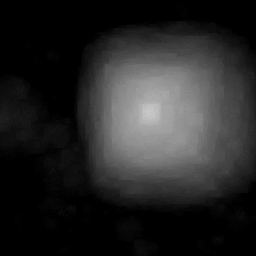}
         \caption{NPI Pos}
         \label{fig:y equals x}
     \end{subfigure}
     \begin{subfigure}[b]{0.16\textwidth}
         \centering
         \includegraphics[width=\textwidth]{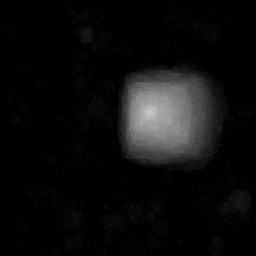}
         \caption{PI Pos-Vel}
         \label{fig:y equals x}
     \end{subfigure}
    \begin{subfigure}[b]{0.16\textwidth}
         \centering
         \includegraphics[width=\textwidth]{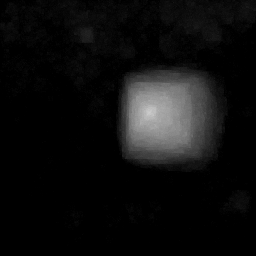}
         \caption{GLR Pos-Vel}
         \label{fig:y equals x}
     \end{subfigure} 
    \begin{subfigure}[b]{0.16\textwidth}
         \centering
         \includegraphics[width=\textwidth]{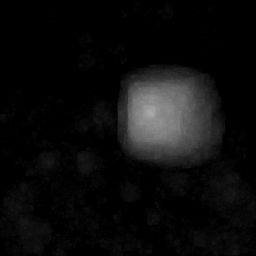}
         \caption{NPI Pos-Vel}
         \label{fig:y equals x}
     \end{subfigure} \\

    \begin{subfigure}[b]{0.16\textwidth}
         \centering
         \includegraphics[width=\textwidth]{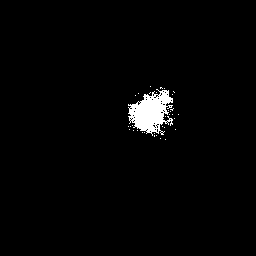}
         \caption{PI Pos}
         \label{fig:y equals x}
     \end{subfigure}
     \begin{subfigure}[b]{0.16\textwidth}
         \centering
         \includegraphics[width=\textwidth]{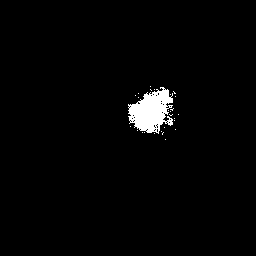}
         \caption{GLR Pos}
         \label{fig:y equals x}
     \end{subfigure}
     \begin{subfigure}[b]{0.16\textwidth}
         \centering
         \includegraphics[width=\textwidth]{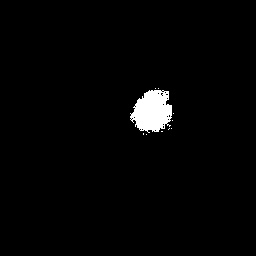}
         \caption{NPI Pos}
         \label{fig:y equals x}
     \end{subfigure}
     \begin{subfigure}[b]{0.16\textwidth}
         \centering
         \includegraphics[width=\textwidth]{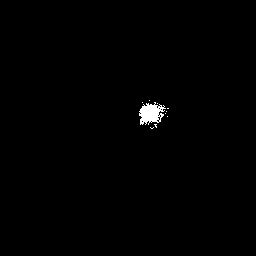}
         \caption{PI Pos-Vel}
         \label{fig:y equals x}
     \end{subfigure}
    \begin{subfigure}[b]{0.16\textwidth}
         \centering
         \includegraphics[width=\textwidth]{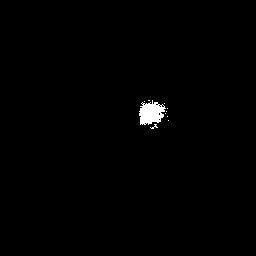}
         \caption{GLR Pos-Vel}
         \label{fig:y equals x}
     \end{subfigure} 
    \begin{subfigure}[b]{0.16\textwidth}
         \centering
         \includegraphics[width=\textwidth]{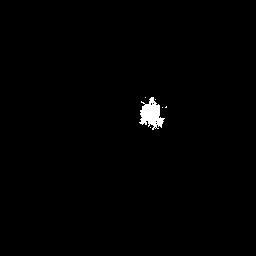}
         \caption{NPI Pos-Vel}
         \label{fig:y equals x}
     \end{subfigure} 
        \caption{Visualization of $H_k(j,t)$ and $H_k(j,t)>\delta$ for various DP-TBD algorithms on real infrared data. Noise power is chosen to yield expected target PSNR of 2.5. (a) is the original uncorrupted frame. (b) is the noisy frame, (c) is the 5,10,20 and 40 pixel radii of the target. (d) is the target trajectory. (e-j) are the direct outputs, $H_k(j,t)$, for the different DP-TBD methods. (k)-(p) are the thresholded outputs, $H_k(j,t)>\delta$, for the different DP-TBD methods. We compare PI, GLR, and NPI for position and position-velocity state spaces.}
\end{figure*}

\begin{table*}[t]
\centering
\small
\begin{tabular}{ |p{1.7cm}||p{1cm}|p{1cm}|p{1cm}|p{1cm}|p{1cm}|  }
 \multicolumn{6}{c}{Table 1: Quantitative Comparison using  Recall and $m$-Precision} \\
  \hline
 Method&   Re & $5$-Pr&  $10$-Pr&  $20$-Pr&  $40$-Pr\\
 \hline
   \multicolumn{6}{|l|}{Position state space with $v_\text{max} = 2$}\\
  \hline
 PI & 0.861 & 0.101 & 0.347 & 0.788 & 0.964\\ 
  GLR & 0.913 & 0.090 & 0.305 & 0.721 & 0.959\\ 
 NPI & 0.878 & 0.093 & 0.315& 0.743 & 0.964\\ 
 \hline
   \multicolumn{6}{|l|}{Position-Velocity state space with $v_\text{max} = 1$, and $a_\text{max} = 1$}\\
  \hline
PI & 0.797 & 0.171 & 0.484 & 0.790 & 0.903\\ 
  GLR & 0.851 & 0.156 & 0.452 & 0.807 & 0.940\\ 
 NPI & 0.721 & 0.217 & 0.517 & 0.796 & 0.917\\ 
 \hline
\end{tabular}
\end{table*}

\subsection{Background Analysis}

In many datasets,  performance is largely  dependent on removing the background signal. Background subtraction refers to removing the background component prior to the weight function, a simplistic method is
\begin{equation}
Z_t^i  \leftarrow  Z_t^{i} - \mu_{Z}^i.
\end{equation}

\noindent where $\mu_Z^i$ is the sample mean of $Z_t^i$ over $t\in \mathcal{T}$. Background subtraction is required for weight functions that rely on the amplitude difference between target and background  state observations (PI, PI-abs, GLR). Edge  normalization refers to removing the background component after the weight function 
\begin{equation}
W_t^{ij}  \leftarrow  \frac{W_t^{ij} - \mu_W^{ij}}{\sigma_W^{ij}}.
\end{equation}

\noindent where $\mu_W^{ij}$ is the sample mean and $\sigma_W^{ij}$ is the sample standard deviation of $W_t^{ij}$ over $t\in \mathcal{T}$. Edge normalization is required for weight functions or data sequences that yield non uniform variance over the state space (NPI).  We use these methods for simplicity; they do not address moving background components, imperfect camera motion alignment, or other complications present in real data. Under such conditions more advanced methods will be necessary.  Errors in removing the background signal are folded into the theory as changing the distributions of target and background edge weights.

\subsection{Complexity Analysis}

The  time complexity to compute the $H_k(j,t)$ for  $|\mathcal{X}| = N$  is $O(kMTN)$ and the space complexity is $O(MTN)$. The actual computation time can be substantially reduced by estimating $H_k(j,t)$ using the previous longest path or by not using a fixed $k$, i.e. $H_t(j,t)$. With the latter there are consequences in the spatial uncertainty, whereas with the former we require $O(kMTN)$ space complexity. To approximate $H_k(j,t+1)$ suppose $\overrightarrow{x} = \{  (x_{t-k},t-k),...,(x_t,t)  \}$ is the longest k-length path to vertex $(x_t,t)$. We estimate 
\begin{equation}
H_{k-1}(x_t,t) \approx    \frac{1}{k-1}  \textstyle\sum_{\tau =t-k+2}^t   W_\tau^{x_{\tau-1},x_\tau},
\end{equation}

\noindent Hence we only need to determine the last edge for the $t+1$ timestep in equation (9). We can repeat this approximation as many times as desired; although the errors will accumulate eventually requiring a full computation of the $H_k(j,t)$ .

\section{Numerical Experiments}

In this section we explore the performance of different DP-TBD algorithms with numerical experiments. Section V.A-D validates the theoretical results from Theorem 2 with a synthetic dataset. Section V.E-H compares different DP-TBD algorithms and state spaces on a real dataset.

\subsection{Synthetic Dataset}

The synthetic dataset experiment is designed to validate the exponential convergence of the error probabilities derived in Theorem 1 and Theorem 2. The dataset consists of simulating a single stationary  point target with amplitude $s = 2.5$, Gaussian noise with standard deviation $\sigma = 1$, and no background signal. These values yield target $\text{SNR} = 2.5$.  The results corresponding to this experiment are in Fig. 3.

\subsection{Synthetic Dataset: Methods}

For this dataset we compare four different DP-TBD methods, PI and PI-abs (Section III.D), GLR (Section III.F), NPI (Section III.G, $\epsilon = 0.01$ and $b = 10^{-5}$), using the position state space (Section III.A, $v_\text{max} = 2$) and the image observation space (Section III.C, $r=0$). For simplicity we use the Chebyshev distance ($L_\infty$ distance) in the state space.  For the decision threshold we use $\lambda = 0.5$ and $0.7$ in  (27).

\subsection{Synthetic Dataset: Evaluation Metrics}

In Fig 3. we numerically compute the error probabilities from Theorem 2 for $\lambda = 0.5$ and $0.7$ as a function of $k$. We measure the false positive states outside a $k$ pixel radius from the target state, $||x_t-y_t||_\infty>k$. This directly corresponds to $\rho = 1/4$ however because the target is stationary it adjusts to  $\rho = 1/2$. Pixel radii are visualized in Fig. 2c and Fig. 4c.

\subsection{Synthetic Dataset: Discussion}

In Fig. 3 all methods exhibit an approximately linear decrease on a logarithmic scale, this corresponds to exponential convergence and is consistent with  Theorem 1 and 2. PI achieves significantly lower error rates than the other methods; it is specifically  tailored to identify positive amplitude targets which in turn results in an inability to detect negative amplitude targets. PI-abs, GLR and NPI, remain applicable with both positive and negative amplitude targets and demonstrate comparable performances. At $k=50$, NPI has lowest false positive and false negative rates followed by PI-abs then GLR. NPI displays a faster rate of convergence as a function of $k$; it has a noticeably steeper asymptotic slope. PI-abs has a slightly better rate of convergence than GLR although the asymptotic slopes of the false positive probability for $\lambda = 0.5$ are nearly identical.

\subsection{Real Dataset}

The real dataset experiment is designed to compare the performance of different DP-TBD algorithms and state spaces in a practical setting. The  SIRSTD (Sequential Infrared Small Target Detection) dataset \cite{10409231} is a real infrared dataset of drones, birds, and other aerial objects that has been annotated with bounding boxes. To make the data more consistent we only select sections where the camera is stationary, downsample large target image sequences by averaging 2x2 pixel blocks, and crop the frames to be 256x256 pixels. This results in 27 image sequences consisting of 250 frames. We add gaussian noise such that  the target peak SNR, averaged over the image sequence, is 2.5.  The  results corresponding to this experiment are in Fig. 4 and Table 1.

\subsection{Real Dataset: Methods}

For this dataset we compare three different DP-TBD methods, PI (Section III.D), GLR (Section III.F), NPI (Section III.G, $\epsilon = 0.01$ and $b = 10^{-5}$), and compare two state spaces, position (Section III.A, $v_\text{max} = 2$), position-velocity (Section III.B, $v_\text{max} = 1$ and $a_\text{max} = 1$) using image observation space (Section III.C, $r=0$). For simplicity we use the Chebyshev distance ($L_\infty$ distance) in both state spaces.  We use $k = 80$ and for the decision threshold we use $\lambda = 0.7$ in (27).

\subsection{Real Dataset: Evaluation Metrics}

In Table 1 we compute the $m$-Precision ($m$-Pr) and Recall (Re) of each construction for $m = 5,10,20$ and $40$. $m$-Pr is the percentage of detected positives within an $m$ pixel radius from the target state. Re is the ratio of true positives to ground truth positives.  $\text{$m$-Pr}= \text{TP}_m/(\text{TP}_m + \text{FP}_m)$  where $\text{TP}_m$ and $\text{FP}_m$ denote the number of detections inside and outside the $m$-radius of a target, respectively. $\text{Re} = \text{TP}/(\text{TP} + \text{FN})$ where TP and FN denote the number true positives and false negatives. We visualize the results of this experiment in Fig. 4.

\subsection{Real Dataset: Discussion}

In Table 1, we observe that performance is broadly consistent across all metrics for methods within the same state space; the major differences occur between the position and position-velocity state spaces. For all weight functions, the position-velocity state space yields a lower recall and lower 40-precision while also a higher 5,10,20-precision. The former is attributable to target motion deviating from the the assumed transition neighborhoods ($v_\text{max} = 1$ and $a_\text{max}=1$) whereas the latter is due to the decreased transition neighborhood size ($M=25$ for position, $M=9$ for position-velocity). NPI displays a greater decrease in recall compared to the other weight functions. This is primarily due to targets being spread over multiple states which causes the normalization in  (22) to reduce the overall signal strength.

Between weight functions, GLR tends to yield higher recall and comparable if not superior performance with 40-precision for both the position and position-velocity state spaces; however PI and NPI achieve higher 5,10-precision. This occurs  because GLR magnifies the effects of SNR: it leverages the high SNR segments to improve tracking during low SNR intervals but this also results in an increased target cloud spread. Some image sequences contain changes in background illumination as a consequence of cloud motion affecting the scene lighting; in these cases NPI shows improved robustness, as the normalization in its construction reduces sensitivity to spatially similar intensity changes. Additionally in a small number of sequences, the target exhibits negative amplitude relative to the background for $\sim 10$ frames, during which PI fails to maintain tracking.

\section{Conclusion and Future Work}

This paper introduces a novel spatial analysis for the DP-TBD class of algorithms specifically designed to address the challenges posed by very low SNR. We establish rigorous theoretical results that demonstrate the inverse relationship between exponential error rate convergence and linear target location uncertainty while also thoroughly detailing the minimum requirements for such methods to be applicable. We numerically validate this result on synthetic data and compare different constructions on a real infrared dataset. We define a new DP-TBD algorithm based on integrating the similarity between observations and demonstrate superior theoretical performance.

The primary application of this theory is in determining regions of interest to gather high quality data and/or apply more computationally intensive algorithms. It can be practically infeasible to collect high resolution data for every region but also computationally infeasible to process all this high resolution data. By using a low resolution and potentially noisy data we can determine a subset of regions to focus data collection and computational resources.

Another immediate application is to combine a method from this framework with one designed to produce precise target masks. The current literature primarily focuses on identifying the exact target mask but under low SNR conditions they are prone to many false positives and fasle negatives. Intersecting the binary outputs with a DP-TBD method may inherit both the precise target mask and minimal false positives.

The first direction for future research is to use NPI to filter background motion. NPI naturally encodes directional motion by using pairs of observations, therefore it can be used to filter out common directions of motion temporally or spatially. An example of common temporal motion would be cars on a highway, they move in specific directions which can be filtered while preserving intersecting orthogonal motion. An example of common spatial motion would be clouds moving in a scene, they all move in similar directions so the local average motion can be filtered out to track contrasting target motion.

The second direction for future research is to use NPI to determine target features. We record the sequence of states that yields the longest path value, then we average the state observations along this path with a $r>0$ in (14). This would give an approximate target appearance which could be used in the construction of  target specific edge weight encodings. NPI is key for this as it inherently retains target identity near complications, like target crossings and moving background elements, by using the relationship between state observations rather than the state observations themselves.

\bibliographystyle{IEEEtran}
\bibliography{bib_fold/sad}

\pagebreak

\section*{Appendix}

\noindent \textit{Lemma 1.1 Proof:} Let $\overrightarrow{y} = \{  (y_{t-k},t-k),...,(y_t,t)  \}$ be the target path and $\overrightarrow{x} = \{  (x_{t-k},t-k),...,(x_t,t)  \}$ be an arbitrary path where $d_\mathcal{G}(x_t,y_t) > m$. First, we show that the last $m_0 = \lfloor \frac{m}{2} \rfloor$ states in any path to $x_t$ cannot be target states, or equivalently $x_\tau \neq y_\tau$ for any $\tau\geq t - m_0$. First note that by the construction, for any path  $\{  (w_t,t),...,(w_{t+\tau},t+\tau)  \}$ in  $\mathcal{G}_\mathcal{T}$, $d_\mathcal{G}(w_{t},w_{t+\tau}) \leq \tau$. Then by the triangle inequality 
\begin{align}
m<  &d_\mathcal{G}(x_t,y_t)\notag\\
 <  &d_\mathcal{G}(x_t,y_{t-m_0}) + d_\mathcal{G}(y_t,y_{t-m_0}) \notag\\
< &d_\mathcal{G}(x_t,y_{t-m_0}) + m/2 \notag 
\end{align}

\noindent  Suppose for contradiction $x_\tau = y_\tau$ for $\tau\geq t - m_0$ then 
\begin{align}
d_\mathcal{G}(x_t,y_{t-m_0}) \leq{}  &d_\mathcal{G}(x_t,x_\tau) + d_\mathcal{G}(y_\tau,y_{t-m_0}) \notag\\
\leq{} &(t-\tau) + (\tau-t-m_0) \notag\\ 
\leq{} & m/2 \notag
\end{align}
This would imply $m < m$, therefore by contradiction $x_\tau \neq y_\tau$ for any $\tau\geq k - m_0$. Then because $\mu_1>\mu_2$ it follows that $\mathbb{E}[W_\tau^{y_{\tau-1}y_\tau}]>\mu_1> \mu_2>\mathbb{E}[W_\tau^{x_{\tau-1}x_\tau}]  $ when $\tau\geq k-m_0$. Additionally $  \mathbb{E}[W_\tau^{y_{\tau-1}y_\tau}] \geq \mathbb{E}[W_\tau^{x_{\tau-1}x_\tau}]$ holds for any $\tau$.  Let $\phi =  \frac{m}{2} - m_0 \in [0,1]$. The path average for $\overrightarrow{y}$ is
\begin{align}
\mathbb{E}[  h_k(\overrightarrow{y})  ]={}  &\mathbb{E}[  \textstyle\frac{1}{k}\textstyle\sum_{\tau=t-k+1}^t  W_\tau^{y_{\tau-1}y_\tau}  ] \notag\\
>{} & \textstyle\frac{1}{k}\textstyle\sum_{\tau=t-k+1}^{t-m_0-1}  \mathbb{E}[  W_\tau^{y_{\tau-1}y_\tau}   ]  \notag \\
& +\textstyle\frac{\phi}{k} \mathbb{E}[  W_{t-m_0}^{y_{t-m_0-1}y_{t-m_0}}   ]  \notag \\
&+\textstyle\frac{1-\phi}{k}\mu_1 +  \frac{m_0-1}{k}\mu_1 \notag \\
>{}&\textstyle\frac{1}{k}\textstyle\sum_{\tau=t-k+1}^{t-m_0-1}  \mathbb{E}[  W_\tau^{y_{\tau-1}y_\tau}   ] \notag\\
& +\textstyle\frac{\phi}{k} \mathbb{E}[  W_{t-m_0}^{y_{t-m_0-1}y_{t-m_0}}   ] +\frac{m}{2k}\mu_1 \notag
\end{align}

\noindent Then using similar logic the path average of $\overrightarrow{x}$ is
\begin{align}
\mathbb{E}[  h_k(\overrightarrow{x})  ]<{}&\textstyle\frac{1}{k}\textstyle\sum_{\tau=t-k+1}^{t-m_0-1}  \mathbb{E}[  W_\tau^{y_{\tau-1}y_\tau}   ] \notag\\
& +\textstyle\frac{\phi}{k} \mathbb{E}[  W_{t-m_0}^{y_{t-m_0-1}y_{t-m_0}}   ] +\frac{m}{2k}\mu_2 \notag
\end{align}
 
\noindent Let $\alpha = \frac{1}{k}\sum_{\tau=t-k+1}^{t-m_0-1}  \mathbb{E}[  W_\tau^{y_{\tau-1}y_\tau}   ] +\frac{\phi}{k} \mathbb{E}[  W_{t-m_0}^{y_{t-m_0-1}y_{t-m_0}}   ]$, therefore we achieve
\begin{align}
\mathbb{E}[  h_k(\overrightarrow{x})  ] < \alpha + \textstyle\frac{m}{2k}\mu_2 \ \text{and} \ \mathbb{E}[  h_k(\overrightarrow{y})  ] > \alpha + \frac{m}{2k}\mu_1.\text{ $\blacksquare$} \notag
\end{align}

\noindent\rule{\linewidth}{0.4pt}\\

\noindent \textit{Lemma 1.2 Proof:}  We note the following
 properties of subgaussian random variables. Let $X$ be $A$-subgaussian and $Y$ be $B$-subgaussian. 
\begin{align}
(1)  \   &\text{\parbox[t]{5.9cm}{If $X,Y$ are independent, then $X+Y$ is $\sqrt{A^2+B^2}$-subgaussian.}}\notag\\
(2)   \   &\text{\parbox[t]{5.9cm}{If $X,Y$ are dependent, then $X+Y$ is $(A+B)$-subgaussian.}}\notag\\
(3)   \   &\text{\parbox[t]{5.9cm}{If $c>0$ then $cX$ is $cA$-subgaussian.}}\notag
\end{align}

\noindent Let $\overrightarrow{x} = \{  (x_{t-k},t-k),...,(x_t,t)  \}$ be an arbitrary path in $G(V,E)$. By construction $W_t^{x_{t-1}x_t} = f(Z_{t-1},Z_t,x_{t-1},x_t)$ which by the HMM means $W_t^{x_{t-1}x_t}$ and $W_{t+\tau}^{x_{t+\tau-1} x_{t+\tau}}$ are independent for $\tau>1$; this is because $Z_t |x_t$ and $Z_{t+\tau}|x_{t+\tau}$ are independent for $\tau\geq 1$. We split the path into two subsets: the even and odd edges. Let $S_e = \{  \tau  :  \tau \in [t-k,t] \ \text{and} \ \tau \ \text{is even}  \}$ and $S_o = \{  \tau  :  \tau \in [t-k,t] \ \text{and} \ \tau \ \text{is odd}  \}$. Because each $W_t^{ij}$ only uses $Z_{t-1}$ and $Z_t$, both
\begin{align}
\{  W_\tau^{x_{\tau-1}x_\tau} - \mathbb{E}[  W_\tau^{x_{\tau-1}x_\tau}  ]   \}_{\tau \in S_e}, \notag\\
\{  W_\tau^{x_{\tau-1}x_\tau} - \mathbb{E}[  W_\tau^{x_{\tau-1}x_\tau}  ]   \}_{\tau \in S_o}, \notag
\end{align}

\noindent are a collection of independent $C$-subgaussians. Then
\begin{align}
 h_k(\overrightarrow{x})- \mathbb{E}[  h_k(\overrightarrow{x})  ]={} &   \textstyle\frac{1}{k}\sum_{i=t-k+1}^t   W_t^{x_{i-1}x_i} \notag\\
 &- \mathbb{E}[  \textstyle\frac{1}{k}\sum_{i=t-k+1}^t     W_t^{x_{i-1}x_i}  ] \notag\\
={}& \textstyle\frac{1}{k}\sum_{\tau \in S_e}   (  W_t^{x_{i-1}x_i} - \mathbb{E}[  W_t^{x_{i-1}x_i}  ]  )\notag\\
&+ \textstyle\frac{1}{k}\sum_{\tau \in S_o}   (  W_t^{x_{i-1}x_i} - \mathbb{E}[  W_t^{x_{i-1}x_i}  ]  )\notag
\end{align}

\noindent From properties (1) and (3),
\begin{align}
\textstyle\frac{1}{k}\sum_{\tau \in S_e}   (  W_t^{x_{i-1}x_i} - \mathbb{E}[  W_t^{x_{i-1}x_i}  ]  ), \notag\\
\textstyle\frac{1}{k}\sum_{\tau \in S_o}   (  W_t^{x_{i-1}x_i} - \mathbb{E}[  W_t^{x_{i-1}x_i}  ]  ), \notag
\end{align}

\noindent are $\sqrt{C^2/2k}$-subgaussian. Therefore from property (2) $  h_k(\overrightarrow{x})- \mathbb{E}[  h_k(\overrightarrow{x})  ] $ is $\sqrt{2C^2/k}$-subgaussian. $\blacksquare$

\noindent\rule{\linewidth}{0.4pt}\\

\noindent \textit{Theorem 1 Proof:} We first derive the relationship  using $d_\mathcal{G}(x_t,y_t)>m$ for $m \in (0,2k]$ then substitute $m = 2\rho k$ for $\rho \in(0,1]$ and any $k\in[\frac{1}{2\rho},\infty)$. We note the following  properties of subgaussian random variables. Let $X_1,...,X_n$ be $C$-subgaussian, not necessarily independent, and $u>0$.
\begin{align}
(1)\  & P( \text{max}_{1\leq i \leq n} \,X_i > u  ) \leq n\exp (  -u^2  /   2C^2  ) \notag\\
(2)\  & P(  X_1 < -u  ) \leq\exp (  -u^2  /  2C^2  ) \notag
\end{align}

\noindent Using (1) we are able to derive another property of subgaussians. We substitute $u = \sqrt{2C^2  \ln(n)} + t$ where $t>0$, then
\begin{align}
P( \text{max}_{1\leq i \leq n}\, X_i > u  ) &\leq n\exp (  -u^2  /   2C^2  )\notag \\
 &\leq \exp (  -(u^2-2C^2\ln(n))  /   2C^2  )\notag \\
   &\leq \exp (  -(t^2 + 2t\sqrt{2C^2  \ln(n)}) /   2C^2  )\notag \\
   &  \leq \exp (  -t^2  /   2C^2  )\notag 
\end{align}

\noindent Which can be summarized as the  following property.
\begin{align}
(3)\  & P( \text{max}_{1\leq i \leq n} \,X_i > \sqrt{2C^2  \ln(n)} + u  ) \leq\exp (  -u^2  /   2C^2  ) \notag
\end{align}

\noindent Now let $\overrightarrow{y} = \{  (y_{t-k},t-k),...,(y_t,t)  \}$ be the target path. By definition $\overrightarrow{y} \in S_k(y_t,t)$ and therefore it follows that $H_k(y_t,t) = \text{max}_{\overrightarrow{x}\in  S_k(y_t,t)} \, h_k(\overrightarrow{x}) \geq h_k(\overrightarrow{y})$. This paired with  Lemma 1.1,  Lemma 1.2, and (3) gives us
\begin{align}
P( H_k(y_t,t) < \delta  )   &\leq   P( h_k(\overrightarrow{y}) < \delta   )\notag\\
   &\leq P( h_k(\overrightarrow{y}) - \mathbb{E}[  h_k(\overrightarrow{y})  ] +\mathbb{E}[  h_k(\overrightarrow{y})  ] < \delta  )\notag\\
      & \leq P(  h_k(\overrightarrow{y}) - \mathbb{E}[  h_k(\overrightarrow{y})  ]  + \alpha + \textstyle\frac{m}{2k}\mu_1 < \delta )\notag\\
   & \leq P(  h_k(\overrightarrow{y}) - \mathbb{E}[  h_k(\overrightarrow{y})  ]  < \delta - \alpha - \textstyle\frac{m}{2k}\mu_1  )\notag\\
   & \leq \exp (  -k( \alpha +  \textstyle\frac{m}{2k}\mu_1 - \delta)^2  / 4C^2  ).\notag
\end{align}

\noindent  when $\alpha + \frac{m}{2k}\mu_1 - \delta > 0$. We define the upper bound for the number of predecessor states as $  |  \mathcal{N}^-_i  |\leq M$. As result, there are at most $M^k$ k-length paths to any given vertex, meaning $|S_k(x_t,t)| \leq M^k$. Additionally, by property of logarithms, $\sqrt{2(2C^2/k)  \ln(M^k)} = \sqrt{4C^2 \ln(M)}$. For simplicity we define this constant as $B = \sqrt{4C^2 \ln(M)}$. Using Lemma 1.1, Lemma 1.2, and (3) 
\begin{align}
P( H_k(x_t,t) > \delta  )   ={}& P(  \text{max}_{\overrightarrow{x}\in S_k(x_t,t)} \, h_k(\overrightarrow{x}) > \delta  )\notag\\
   \leq{}&P(  \text{max}_{\overrightarrow{x}\in S_k(x_t,t)} \, (  h_k(\overrightarrow{x}) - \mathbb{E}[  h_k(\overrightarrow{x})  ]\notag\\
   & \ \ \ \ + \mathbb{E}[  h_k(\overrightarrow{x})  ]  ) > \delta  )\notag\\
   \leq{}&P(  \text{max}_{\overrightarrow{x}\in S_k(x_t,t)} \, (  h_k(\overrightarrow{x}) - \mathbb{E}[  h_k(\overrightarrow{x})  ] \notag\\
   & \ \ \ \ + \alpha + \textstyle\frac{m}{2k}\mu_2  ) > \delta  )\notag\\   
   \leq{}&P(  \text{max}_{\overrightarrow{x}\in S_k(x_t,t)} \, (  h_k(\overrightarrow{x}) - \mathbb{E}[  h_k(\overrightarrow{x})  ]   ) \notag\\
   & \ \ \ \ > \delta - \alpha - \textstyle\frac{m}{2k}\mu_2    )\notag\\  
   \leq{}&P(  \text{max}_{\overrightarrow{x}\in S_k(x_t,t)} \, (  h_k(\overrightarrow{x}) - \mathbb{E}[  h_k(\overrightarrow{x})  ]   )  \notag\\
   & \ \ \ \ >   B + \delta - \alpha - \textstyle\frac{m}{2k}\mu_2- B\notag\\ 
   \leq{}&\exp (  -k( \delta - \alpha - \textstyle\frac{m}{2k}\mu_2 - B)^2  /  4C^2  )\notag
\end{align}

\noindent when $\delta - \alpha - \frac{m}{2k}\mu_2 - B > 0$. To reduce the number of variables we set these error rates to be equal. This results in the  following statements
\begin{align}
(4) \ & \alpha + \textstyle\frac{m}{2k}\mu_1 - \delta = \delta - \alpha - \textstyle\frac{m}{2k}\mu_2 - B\notag\\
(5) \ & \alpha + \textstyle\frac{m}{2k}\mu_1 - \delta > 0\notag\\
(6) \ & \delta - \alpha - \textstyle\frac{m}{2k}\mu_2 - B > 0\notag
\end{align}

\noindent Adding (5) and (6) yields the constraint that $ \ \frac{m}{2k}(\mu_1-\mu_2) > B$. Solving (4) for $\delta$ yields
\begin{align}
\delta &  = \alpha + \textstyle\frac{1}{2}(  \frac{m}{2k}\mu_1 + \frac{m}{2k}\mu_2 + B  )\notag\\
 & =   \textstyle\frac{1}{2}(  \alpha + \frac{m}{2k}\mu_1  ) + \frac{1}{2}(  \alpha + \frac{m}{2k}\mu_2 + B  ).\notag
\end{align}

\noindent Substituting $\delta$ into our previous inequalities gives us
\begin{align}
\textstyle P(  H_k(y_t,t) < \delta  ) \leq \exp (  -k(  \frac{m}{2k}(\mu_1 - \mu_2) - B )^2  /  16C^2  )\notag\\
\textstyle P(  H_k(x_t,t) > \delta  ) \leq \exp (  -k(  \frac{m}{2k}(\mu_1 - \mu_2) - B )^2  /  16C^2  )\notag
\end{align}

\noindent when $  \frac{m}{2k}(\mu_1-\mu_2)  > B$. Now set $m = 2\rho k$ for a chosen $\rho \in (0,1]$. If $  \rho(\mu_1-\mu_2)  > \sqrt{4C^2 \ln(M)}$  then  there exists a $A = (  \rho(\mu_1 - \mu_2) - \sqrt{4C^2 \ln(M)} )^2  / 16C^2>0$ independent of $k$ and $t$ and a $\delta = \alpha + (  \rho\mu_1 + \rho\mu_2 + \sqrt{4C^2 \ln(M)}  )/2$ such that for any non-target state $x_t$ satisfying $d_\mathcal{G}(x_t,y_t)>2\rho k$
\begin{align}
\textstyle P(  H_k(y_t,t) < \delta  ) &< \exp (  -Ak  )\notag,\\
\textstyle P(  H_k(x_t,t) > \delta  ) &< \exp (  -Ak  ).\text{ $\blacksquare$}\notag
\end{align}

\noindent\rule{\linewidth}{0.4pt}\\

\noindent \textit{Lemma 2.1 Proof:} We note the following
 properties of subgaussian random variables. Let $X_1,...,X_n$ be $C$-subgaussian, not necessarily independent.
\begin{align}
     (1)\  \mathbb{E}[ \text{max}_{1\leq i \leq n}\, X_i ] \leq \sqrt{2C^2 \ln(n)} \notag
\end{align}

\noindent Let $\overrightarrow{x}\in S_k(x_t,t)$  where $d_\mathcal{G}(x_t,y_t)>2k$ from Lemma 1.1, $\mathbb{E}[  h_k(\overrightarrow{x})   ] <\mu_2 < \alpha+\rho \mu_2$ for $a$  with $m = 2\rho k$. Additionally by assumption $\rho(\mu_1-\mu_2)>\sqrt{4C^2\ln(M)}$, so
\begin{align}
   \mathbb{E}[  H_k(x_t,t)   ]<{}& \mathbb{E}[  \text{max}_{\overrightarrow{x}\in S_k(x_t,t)} \, ( h_k(\overrightarrow{x}) - \mathbb{E}[  h_k(\overrightarrow{x})  ]  )    ] \notag\\
   & + \alpha + \textstyle\rho\mu_2 \notag\\
   <{}& \textstyle\sqrt{4C^2 \ln(M)} + a + \rho\mu_2\notag\\
   <{}& \textstyle \alpha + \frac{1}{2}( \rho\mu_1 + \rho\mu_2 + \sqrt{4C^2 \ln(M)}  )\notag\\
   ={} & \delta \notag
\end{align}

\noindent In a similar manner
\begin{align}
   \mathbb{E}[  H_k(y_t,t)   ]\geq{}& \mathbb{E}[  h_k(\overrightarrow{y})   ] \notag\\
   >{}& \textstyle \alpha + \rho\mu_1\notag\\
   >{}& \textstyle \alpha + \frac{1}{2}( \rho\mu_1 + \rho\mu_2 + \sqrt{4C^2 \ln(M)}  )\notag\\
   ={} & \delta \notag
\end{align}

\noindent Therefore $\mathbb{E}[  H_k(x_t,t)   ] < \delta < \mathbb{E}[  H_k(y_t,t)   ]$ so there exists an $\lambda$ such that
\begin{align}
\delta =  \lambda\mathbb{E}[  H_k(y_t,t)   ] + (1-\lambda)\mathbb{E}[  H_k(x_t,t)   ].\text{ $\blacksquare$} \notag
\end{align}

\noindent\rule{\linewidth}{0.4pt}\\

\noindent \textit{Theorem 2 Proof:} First, fix $k\in [\frac{1}{2\rho_1},\infty)$ and $t\in [k,\infty)$. Then let $\rho_0 =  \sqrt{4C^2 \log(M)}/(\mu_1 - \mu_2)$ and let $A_0$ and $A_1$ be the  convergence rates for $\rho_0$ and 1. Define $\lambda(\rho,k,t)$ as the mixing coefficient in Lemma 2.1 for $\rho,k,t$, where  
\begin{subequations}
\begin{align}
\beta_1 &= \text{min}_{t\in[1,\infty)} \, \text{min}_{k\in[1,\infty)} \, \lambda(\rho_0,k,t),\\
\beta_0 &= \text{max}_{t\in[1,\infty)} \, \text{max}_{k\in[1,\infty)} \, \lambda(1,k,t).
\end{align}
\end{subequations}

\noindent We require that $\beta_1\geq \beta_0$; for constant SNR targets with no background, if the applicability criteria is satisfied then this will always hold where $\beta_1 \approx 1$ and $\beta_0 \approx 1/2$. In practice we assume it to be true and it can be influenced by upper bounding the edge weights. Now define the functions
\begin{align}
 f_1(\rho)   &= \textstyle \alpha + \frac{1}{2}(  \rho\mu_1 + \rho\mu_2 + \sqrt{4C^2  \ln(M)}  )\notag\\
f_2(\rho)   &= \textstyle (  \rho(\mu_1 - \mu_2) - \sqrt{4C^2 \ln(M)} )^2  / 16C^2\notag\\
f_3(\lambda) &=  \lambda\mathbb{E}[  H_k(y_t,t)   ] + (1-\lambda)\mathbb{E}[  H_k(x_t,t)   ]\notag
\end{align}

\noindent  $f_1(\rho)$ is a continuous strictly monotonically decreasing function, and  $f_2(\rho)$ is a continuous strictly monotonically decreasing function for $\rho\in [\rho_0,1]$. The former is because  $\alpha= (1-\rho)(\mu_1+\epsilon(\rho))$ where $\epsilon(\rho)\geq0$ is a decreasing piecewise constant fucntion (sum of the differences between $\mu_1$ and the actual expected edge weight for the first $(1-\rho) k$ edges). $f_3(\lambda)$ is a continuous strictly monotonically increasing function when $\lambda\in [0,1]$.  By properties of monotonic functions  $f_1^{-1}(f_3(\lambda))$ and  $f_2(f_1^{-1}(f_3(\lambda)))$ are monotonically decreasing,  for $\lambda \in [f_3^{-1}(f_1(1)),f_3^{-1}(f_1(\rho_0))] = [\lambda_0,\lambda_1]$. Then, 
\begin{align}
 f_3^{-1}(f_1(\rho_0))& = \lambda(\rho_0,k,t) \geq   \beta_1,\notag\\
 f_3^{-1}(f_1(1))&= \lambda(1,k,t) \leq \beta_0,\notag
\end{align}

\noindent so $[\beta_0,\beta_1] \subseteq [\lambda_0,\lambda_1]$ for any $k,t$. $f_1(\rho)$ and $f_2(\rho)$ are the threshold and convergence rate for $\rho,k,t$ so $f_2(f_1^{-1}(f_3(\lambda)))$  and $f_1^{-1}(f_3(\lambda))$ are $A$ and $\rho$ for $\lambda,k,t$. Now let $\lambda \in [\beta_0,\beta_1]$, 
\begin{align}
f_2(f_1^{-1}(f_3(\lambda)))    & \geq f_2(f_1^{-1}(f_3(f_3^{-1}(f_1(\rho_0)))))\notag \\
    & \geq f_2(\rho_0)\notag \\
f_1^{-1}(f_3(\lambda)) & \leq f_1^{-1}(f_3(f_3^{-1}(f_1(1))))\notag\\
   & \leq 1\notag
\end{align}

\noindent  Therefore for any  $\lambda \in[\beta_0,\beta_1]$ there exists $A\in (0,\infty)$ and $\rho \in (0,1]$, both independent of $k$ and $t$, such that for any non-target state $x_t$ satisfying $d_\mathcal{G}(x_t,y_t)>2\rho k$
\begin{align}
P(  H_k(y_t,t)   <   \delta  )   &<   \exp(  -Ak  ),\notag\\
P(  H_k(x_t,t)   >   \delta  )   &<  \exp( -Ak  ). \text{ $\blacksquare$}\notag
\end{align}

\end{document}